# Optic Detectors Calibration for Measuring Ultra-High Energy Extensive Air Showers Cherenkov Radiation by 532 nm Laser

S. P. Knurenko[1, a], Yu. A. Egorov[1, b], I. S. Petrov[1, c]

Yu.G. Shafer Institute of Cosmophysical Research and Aeronomy
SB RAS, 31 Lenin Ave., 677980 Yakutsk, Russia.

**ABSTRACT**

Calibration of a PMT matrix is crucial for the treatment of the data obtained with Cherenkov tracking detector. Furthermore, due to high variability of the aerosol abundance in the atmosphere depending on season, weather etc. A constant monitoring of the atmospheric transparency is required during the measurements. For this purpose, besides traditional methods, a station for laser atmospheric probing is used.

## 1. INTRODUCTION

At the Yakutsk array, atmosphere laser sensing station is deployed. Laser radiation of remote atmosphere sensing attenuates when passing through molecules of gases and aerosols of the atmosphere. Part of radiation scatters off aerosols particles and falls through a narrow gap on tiled PMT. Amplitude of the signal determines by physics and properties of atmosphere to scatter photons of Cherenkov light. In our case, analysis of these data allows to clarify some air showers parameters, which are determined from Cherenkov radiation, for example, estimation of shower energy [1], determined as all ionization losses of electrons and muons in the atmosphere. In addition, LIDAR allows to determined spectral transparency of atmosphere directly [2], with respect to previously used in [3] relative method.

At Yakutsk array, measurement of Cherenkov radiation characteristics carried out by large number of integral and track detectors [4]. Cherenkov detectors operate only during moonless and clear nights. Track detector consists of chamber with narrow straight and long gap under which located perpendicularly to gap direction tiles of PMT (Fig. 1). Each PMT of track detector registers Cherenkov photons throughout dense part of the atmosphere. Such detector can trace the entire longitudinal development of the shower particles and allows us to study the physics of the process. Since these measurements are precise nature, the accuracy of the cascade curve is crucial for this technique, LIDAR data can significantly improve the accuracy of the experiment.

---

[a]s.p.knurenko@ikfia.sbras.ru, [b]zeppelin@ikfia.sbras.ru, [c]igor.petrov@ikfia.sbras.ru

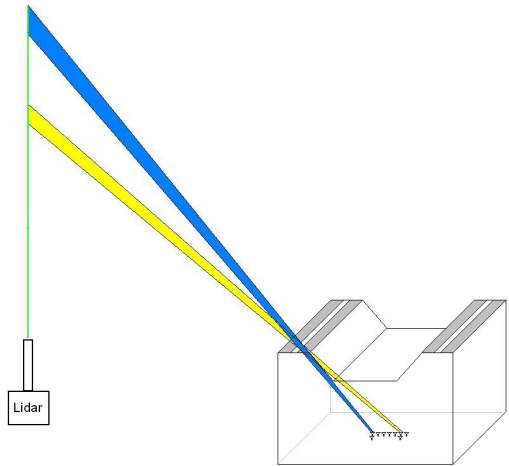

Fig. 1. Scheme of calibration of Cherenkov track detector with 532 nm laser.

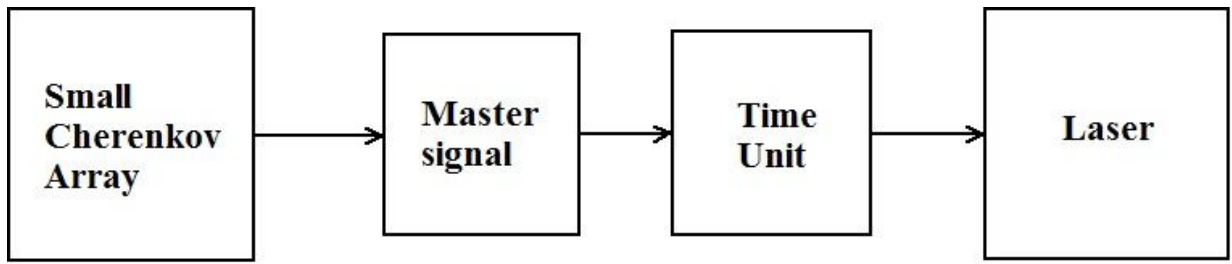

Fig. 2. Laser launch control flowchart for calibration of Cherenkov detectors.

## 2. WEATHER CONDITION REQUIREMENTS FOR CHERENKOV OBSERVATIONS

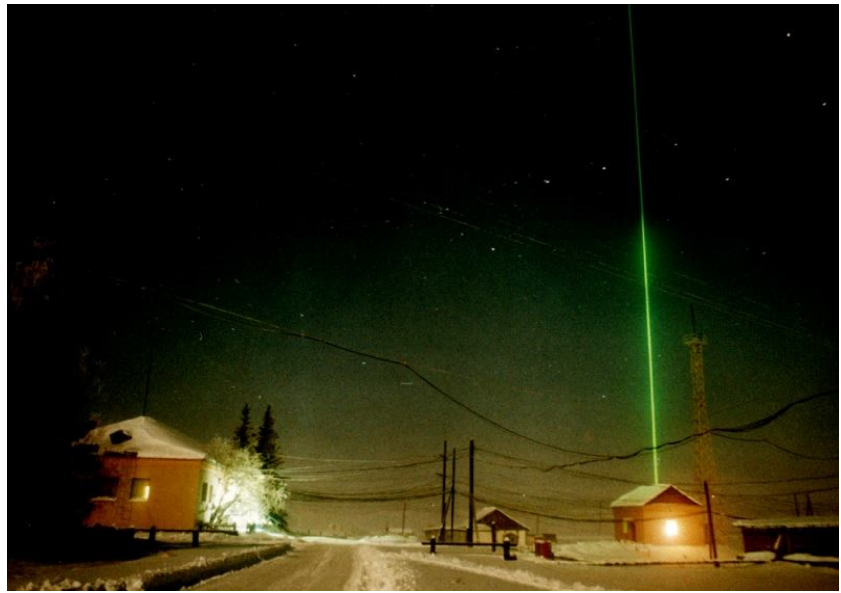

Fig. 3. LIDAR beam at Yakutsk array

Cherenkov radiation observation conducted mainly in the wintertime (for Yakutia, October - April), when the atmosphere is practically frozen out and the sky partly cloudy or clear [2]. In the absence of man-made and the Moon lights during this period has a minimum background light of the night sky in the wavelength range from 300 nm to 800 nm, which is the ideal condition for optical observations. However, as long-term observation implemented various weather conditions, the regular measurements of atmospheric transparency needed. In our case, they do contribute to clarifying estimates of various characteristics of EAS in the calculations by using the correction factors, taking into account the transparency of the atmosphere [6].

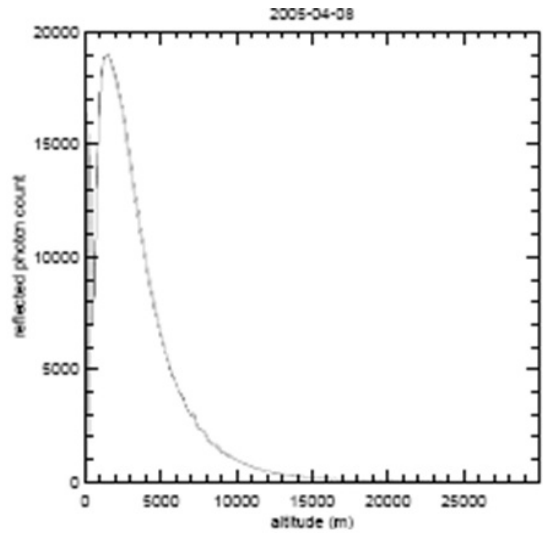

Fig. 4. LIDAR measurements of transparency of the atmosphere

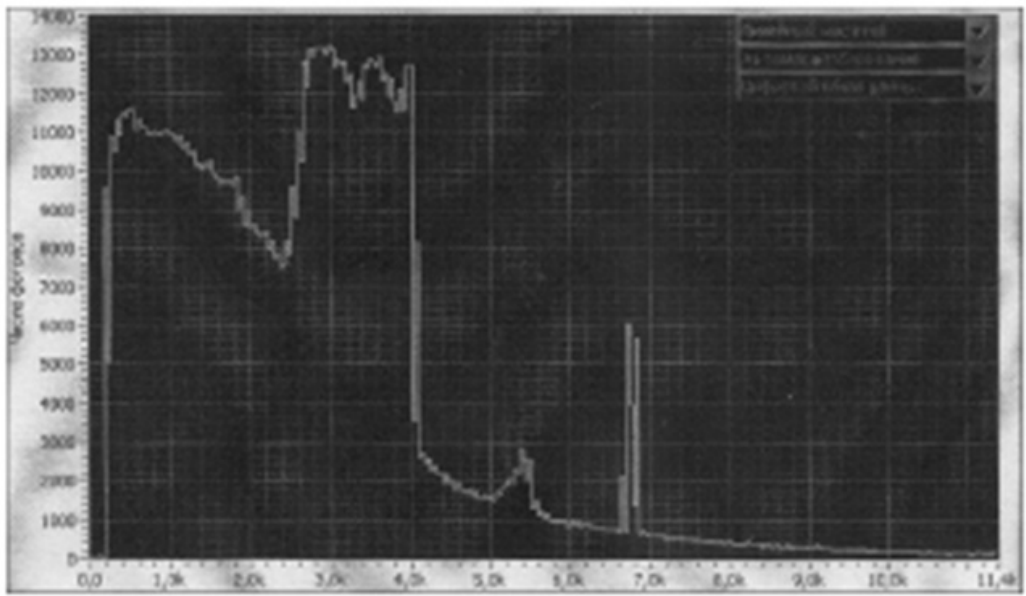

Fig. 5. LIDAR measurements of transparency of the atmosphere (low height).

Cherenkov light observations requires an information about transmission of the atmosphere at the time of observation. Therefore, an essential part of these observations is to organize regular measurements of the spectral or integrated transparency of the atmosphere. There are many ways of measuring the transparency of the atmosphere [1, 2, 3, 4, 5], but the optimal method is based on measurement of the same Cherenkov radiation [4, 5]. The LIDAR measurements is also suitable as monitor of the state of the atmosphere, especially when used for this purpose tunable to different wavelength. The Yakutsk array uses laser at a wavelength of 532 nm, which is close to the maximum of the spectral characteristics of the PMT-49B, which is used in the Cherenkov detector as a receiver of EAS Cherenkov light.

Profile measurements of the transmission function of the atmosphere conducted in the area of Yakutsk array showed that in periods of observation implemented a number of scenarios of atmospheric conditions, from Rayleigh scattering, when practically note the clear atmosphere (Fig. 4), to almost full or partial absorption of laser radiation in a thick continuous layer of clouds or to separate layers, as shown in the Figure 5.

## 3. CALIBRATION ALGORITHM CHERENKOV DETECTORS USING LASER

Since the spectral characteristics of Cherenkov detectors is within wavelength range 300-800 nm, selected monochromatic laser λ = 532 nm with a sufficiently high output power is able to create a flash of light up to stratospheric heights. Which is more than enough to overlap the maximum luminescence of air shower Cherenkov light (maximum is located at a height of 3 - 5 km above sea level), which allow us to observe photons from both initial stage of shower development and from lower layers of the atmosphere near sea level. The latter allow us to control the sensitivity of the detector near threshold level and use the signals as a test measurement.

The response time of LIDAR is constant and has an exponential shape with the only Rayleigh scattering and may be used for calibration, because area under the curve reflects total number of photons collected from all heights. Comparison of LIDAR response and PMT response at the same solid angles by comparing areas under the curves and allows one to calibrate the track detector.

For differential detectors, which overlook certain height, this fact makes it possible to directly attribute to the response of the PMT, i.e. pulse area known number of photons generated by the laser (see Fig. 1).

Table 1. Laser radiation heights visible by track detector PMT.

| № PMT | 1 | 2 | 3 | 4 | 5 | 6 | 7 | 8 | 9 | 10 |
|---|---|---|---|---|---|---|---|---|---|---|
| Min Height, m | 1134 | 972 | 852 | 756 | 681 | 618 | 567 | 525 | 1047 | 546 |
| Max Height, m | 1455 | 1197 | 1020 | 885 | 885 | 702 | 639 | 582 | 1314 | 609 |

We restored height of each monitored PMT (Table) by using geometric method and taking into account the aperture of Cherenkov detectors and pulse parameters including pulse area are assigned for these heights. In this case, the pulse reflects the number of photons that come from specified heights and with certain intensity. Changes of intensity and hence the pulse form is associated with only optical signal propagation conditions in the atmosphere and therefore with good weather conditions (only in the case of Rayleigh scattering), this signal can be considered as calibration signal.

## 4. METHOD OF CALIBRATION AND MEASUREMENTS TESTING OF EAS CHERENKOV LIGHT

Figure 2 shows a flowchart of process control "Shooting" laser, permanently installed in the center of Yakutsk array and "shoot" vertically to the plane of device. The control was carried out by Cherenkov trigger of Small Cherenkov array after the laser power ups at predetermined wavelength λ=532 nm.

Besides from being used as calibration for Cherenkov detectors, the laser used as a test of the optical part of the Yakutsk array and for measurement of atmosphere transparency. The "shooting" session occurred after each event of air showers or every 15 minutes. As the station of LIDAR measurements is located at the center of the array (Fig. 6), then we can calibrate Cherenkov detectors without change of the output characteristics of the laser.

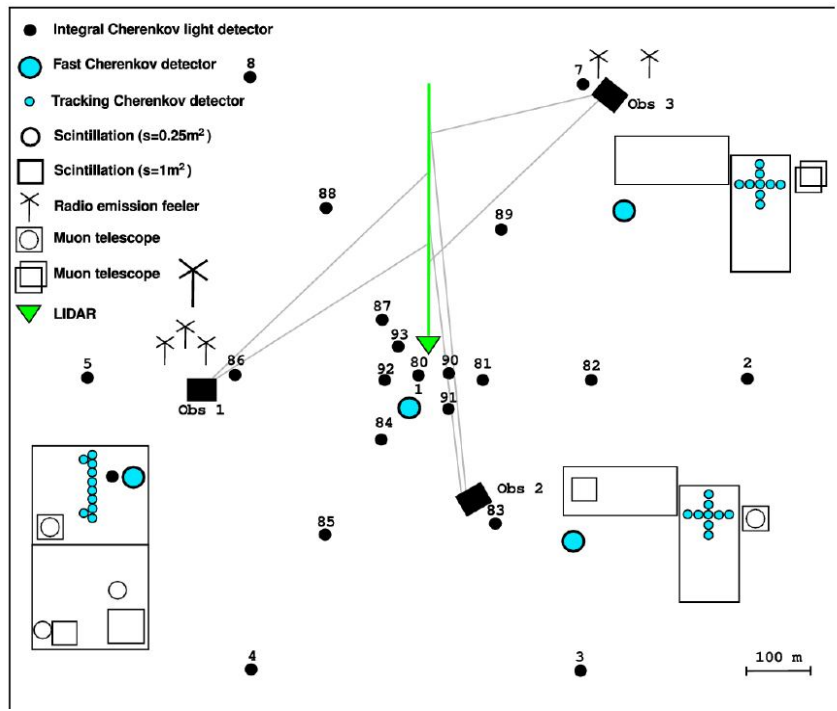

Fig. 6. Small Cherenkov array.

## 5. CONCLUSION

Laser testing in 2013 season showed a sufficiently high stability of the signal reproduction of track detectors up to 20 Hz. Registered pulses had a constant amplitude and a constant shape that is necessary for calibration.

Statistical analysis showed that the accuracy of counting the number of photons for each PMT in the case of vertical sounding and great weather conditions does not go beyond 15 ± 3%. This is significantly higher than if the Cerenkov receivers calibrated using a reference block of methyl methacrylate [5].

**REFERENCES**

[1] Knurenko, S. P., Ivanov, A. A., Sleptsov I .Ye., Sabourov A. V., JETP Let., 2006, V. 83, 11, p. 563-567 (in Russian).
[2] Knurenko S.P., Nikolashkin S.V., Saburov A.V., Sleptsov I.Ye., Study of atmosphere characteristics using ultra-high energy cosmic ray and $\lambda$ = 532 nm LIDAR. // Proc. Of SPIE. Vol. 6522 (2006).
[3] Dyakonov M. N., Knurenko S. P., Kolosov V. A., Sleptsov I. Ye., Optics of atmosphere, 1991, V. 4, p. 868 – 873 (in Russian).
[4] Knurenko S.P., Sabourov A.V., The Depth of Maximum Shower Development and Its Fluctuations: Cosmic Ray Mass Composition at $E_0 \geq 10^{17}$ eV. // Astrophys. Space Sci. Trans., 7, 251-255, 2011. www.astrophys-space-sci-trans.net/7/251/2011/ doi:10.5194/astra-7-251-2011.
[5] Ivanov A.A., Knurenko S.P., Sleptsov I.Ye., Measuring extensive air showers with Cherenkov light detectors of the Yakutsk array: the energy spectrum of cosmic rays. New J. Phys. 11. No 6 (June 2009), 065008 (30 pp).
[6] D'yakonov M. N., Knurenko S. P., Kolosov V. A., Sleptsov I. Ye., Reconstruction of the vertical profile of mean atmospheric transmittance from data of optical observations of cosmic rays. Optic of atmosphere and ocean, 1999, 12, 4, 315-320.